\documentclass[aps,prb,twocolumn]{revtex4}
\usepackage{tabularx,graphicx}
\usepackage{amsmath, amsthm, amssymb} 
\usepackage{bbm}

\def\be{\begin{equation}}
\def\ee{\end{equation}}
\def\bea{\begin{eqnarray}}
\def\eea{\end{eqnarray}}
\def\ba{\begin{array}}
\def\ea{\end{array}}

\newcommand{\vk}{{\bf {k}}}

\newcommand{\vkq}{{\bf {k}}-{\bf{q}}}

\begin{document}

\title{Quantum-Critical transport at a semimetal-to-insulator transition on the honeycomb lattice }

\author{Lars Fritz}
\affiliation{Institut f\"ur Theoretische Physik,
Universit\"at K\"oln, Z\"ulpicher Stra\ss e 77, 50937 K\"oln, Germany}

\date{\today}

\begin{abstract}
In this paper we study transport properties of electrons on the two-dimensional honeycomb lattice. We consider a half-filled system in the vicinity of a symmetry-breaking transition from a semimetallic  phase towards an insulating phase with either charge density or spin density wave order. The effect of either order is to break the sublattice inversion symmetry which induces a finite gap for the electronic single-particle excitations. Phenomenologically, such a scenario is described in the framework of a Gross-Neveu theory. We analyze two related formulations of the model by means of (i) a controlled renormalization group calculation and (ii) the large-N method, both of which in combination with a Boltzmann transport equation. We determine the quantum-critical conductivity and also discuss crossover behavior from quantum critical behavior into the insulating and/or the semimetallic phases. We find that at asymptotically low temperatures the quantum-critical conductivity is given by a temperature independent universal number. Over a large temperature window the temperature independent quantum critical conductivity is masked by a logarithmically temperature dependent contribution due to the marginally irrelevant long-range Coulomb interaction. We discuss possible origins of this peculiarity in the two complementary formulations of the model. Furthermore, we consider possible relations of our findings to recent experiments, with a special emphasis on the quantum-critical-to-insulator crossover. We find that our results are in remarkably good qualitative and quantitative agreement with a recent analysis of the data sets under the hypothesis of an underlying gap in the single-particle spectrum.
\end{abstract}

\maketitle


\maketitle

\section{Introduction}\label{intro}

Two dimensional electronic systems on a honeycomb lattice have seen tremendous activity over the last couple of years. This development was triggered by the experimental isolation of single layer graphene which comes with a variety of fascinating and unusual properties~\cite{NetoRMP}. 

Commonly, graphene is considered an excellent conductor with extremely high mobilities~\cite{stormer}. The effective low-energy theory describing the system is given in terms of massless Dirac fermions moving at the Fermi velocity $v_F$ instead of the speed of light $c$ ($\frac{v_F}{c}\approx \frac{1}{300}$). Most of the experimental findings in graphene to date can be explained in terms of weakly interacting massless Dirac quasiparticles~\cite{Zhou06}. A most prominent exception is provided by the recently observed fractional quantum Hall effect~\cite{Du, Abanin,Bolotin}.

Despite this predominant lack of signatures of correlation effects, many authors have analyzed the role of local and long-range Coulomb interaction and possible interaction driven insulating phases~\cite{Semenoff,Herbut3,Herbut2,Herbut1,Sorella,Martelo,Paiva,Gorbar,Khevshchenko,Drut,Vafek,Zheng}.

In this paper we study transport in the vicinity of a semimetal-to-insulator quantum phase transition, in which the opening of the single-particle gap in the insulating phase is driven by spontaneous sublattice symmetry breaking. The sublattice inversion symmetry breaking is driven by collective instabilities of the charge density or spin density wave type. Note that this scenario does not describe the Mott transition in the traditional sense due to the additional symmetry breaking~\cite{Peierls}. In that sense the treatment presented here is closer to the excitonic insulator obtained due to long-range Coulomb interaction and chiral symmetry breaking~\cite{Khevshchenko,Drut} than it is to a pure Hubbard model. Within this paper we analyze the simplest possible transition from a semimetal to an insulator described by a Hubbard like model and discard the possible existence of interesting intermediate phases such as a spin liquid phase~\cite{Zheng}.

Transport properties at quantum-criticality have traditionally been a very active branch of research in strongly correlated matter. In two dimensions microscopic calculations have been performed with the quantum-critical conductivity being given by a temperature independent universal number~\cite{ssbook,Damle,ssqhe,bhaseen}.  This has to be expected for dimensional reasons if the critical point of the field theory has relativistic invariance and is located at finite interaction strength. 

In the context of clean intrinsic graphene, which shares a number of characteristics with a quantum-critical system~\cite{Sheehy}, the critical conductivity was shown to diverge in logarithmic manner upon lowering temperature~\cite{FSMS}. This behavior is rooted in the renormalization group fixed point of the system lying at zero interaction, with the interaction parameter describing long-range Coulomb interaction being marginally irrelevant, thus flowing to zero logarithmically upon lowering the energy scale.

In our problem, the insulating phase is characterized by the presence of collective order of charge density (CDW) or spin density wave (SDW) type~\cite{Herbut1}. The finite order parameter acts like a mass term for the Dirac fermions and consequently opens a gap in the electronic excitation spectrum. Both sorts of long-range order break sublattice inversion ($Z_2$) symmetry (conventionally referred to as chiral symmetry), which is the driving force behind the finite electronic gap. SDW additionally breaks spin rotation symmetry (SU($2$)).

Within this paper we choose a simple field theoretic formulation of the Gross-Neveu type~\cite{GrossNeveu,ZinnJustin}, which captures the collective instability and the opening of the electronic quasiparticle gap. We consider two related formulations: (i) a Landau-Ginzburg type bosonic order parameter theory coupled to Dirac fermions via a Yukawa-type coupling (henceforth referred to as model I) and (ii) a locally interacting theory of Dirac fermions (henceforth referred to as model II). In both cases there is a quantum phase transition of the semimetal-to-insulator type, which is second order and described by an interacting fixed point. In both cases an appropriate order parameter can easily be defined, which is zero in the semimetallic phase and finite in the insulating phase. Thinking in terms of the Landau-Ginzburg description of phase transitions, the order parameter symmetry is $O(N)$ with $N=1$ (Ising or $Z_2$) for CDW and $N=3$ (Heisenberg) for SDW.

\begin{figure}[t]
\includegraphics[width=0.45\textwidth]{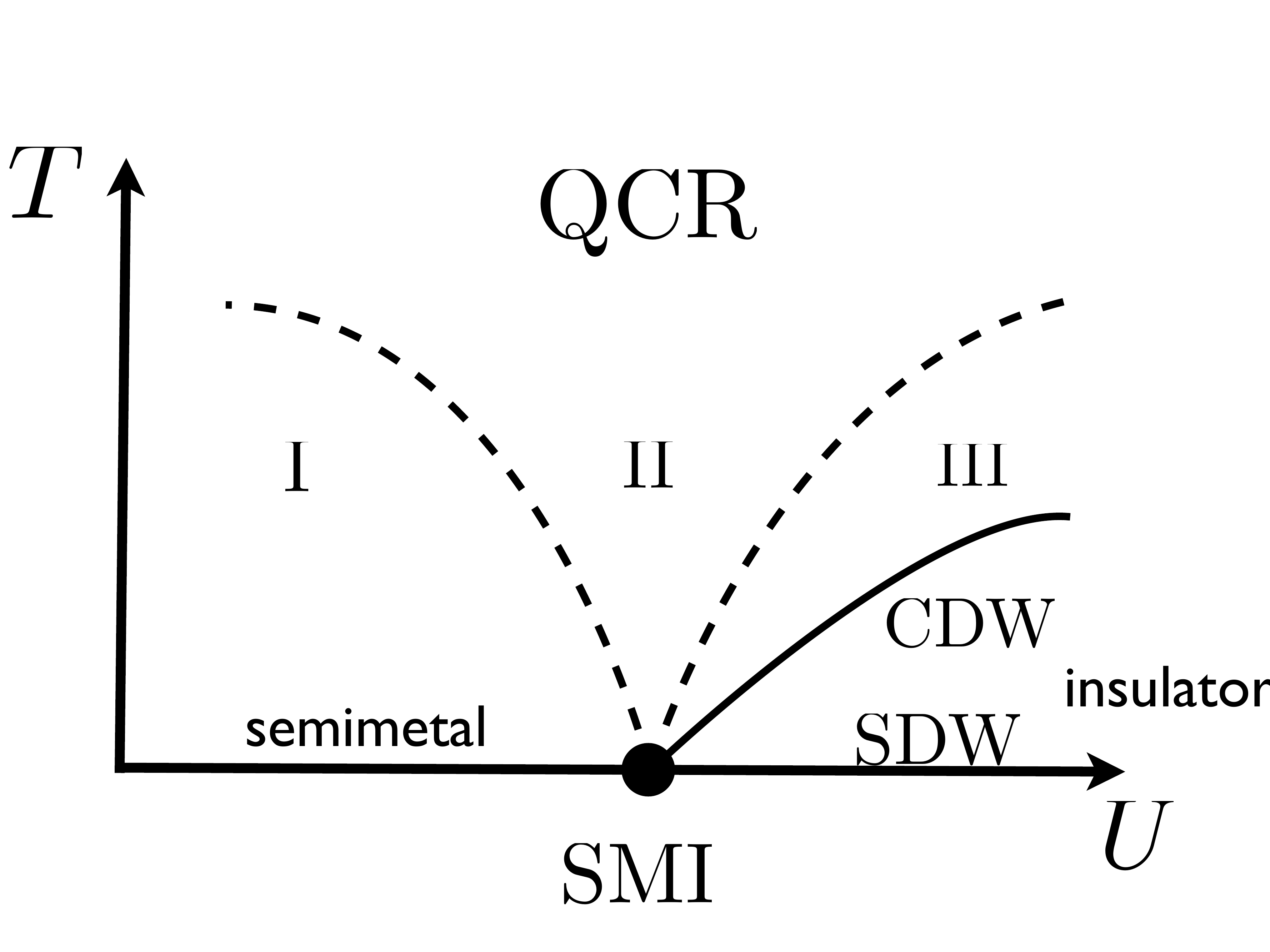}
\caption{Schematic phase diagram. The dashed lines denote crossover lines, whereas full lines denote phase transitions. The different regions and the crossovers are discussed in the main text. The insulating region III differs for the scenarios CDW and SDW. Whereas for CDW there is a finite temperature phase transition of the Ising type, there is no finite temperature transition for the SDW due to the Hohenberg-Mermin-Wagner theorem with forbids the breaking of a continuous symmetry at finite temperature in two dimensions.}\label{Fig:phasediag}
\end{figure}

Microscopically, local interactions can trigger these instabilities: on-site Coulomb repulsion favors SDW, whereas repulsion between adjacent sites favors CDW. Fig.~\ref{Fig:phasediag} sketches a phase diagram with the vertical axis being temperature and the horizontal axis the local interaction parameter. The strength of local repulsion, denoted $U$, serves as tuning parameter for the quantum phase transition. In the case of a quantum phase transition towards CDW, $U$ denotes repulsive interaction between adjacent sites, whereas for the case of a transition towards SDW the repulsion is on-site. Our analysis proceeds along the lines of a combination of a semiclassical Boltzmann equation combined with renormalization group arguments (model I) and a large-N expansion (model II)~\cite{ssbook,Damle,ssqhe,bhaseen,FSMS,MFS,MSF,Kashuba}. Throughout the paper all mathematical expressions are explicitly shown for the CDW case. The SDW expressions differ in combinatorial factors due to the difference in order parameter symmetry, but not in their structure. However, qualitative differences occur at finite temperature due to the absence of long-range order in the SDW case. In the CDW case a finite gap in the electronic excitation spectrum is stable at finite temperatures, whereas there is no hard gap for the SDW. This is discussed in great detail in the conclusions.

\subsection{Overview of the results}
We calculated the quantum-critical minimal conductivity of graphene at the semimetal-to-insulator transition using two different phenomenological models. Furthermore, we estimated crossover functions for the conductivity which obtain upon lowering temperature and entering the semimetal or respectively the insulating phase. Most interestingly, we estimate the crossover function for the quantum-critical-to-insulator crossover and compare it to experimental data in graphene. 

Considering two slightly distinct models we find seemingly contrasting transport properties:

(1) Within model I in contrast to other relativistically invariant critical points at finite interaction strength in two dimensional systems, the d.c. conductivity is not independent of temperature, but instead seemingly diverges upon lowering temperature~\cite{ssbook,Damle,ssqhe,bhaseen}. This surprising result obtains because of a conspiracy of matrix elements and kinematical constraints which prohibits electronic current relaxation from the Yukawa coupling to {\it all} orders in perturbation theory. This statement, however, is only strictly true if the electrons and bosons are treated as sharp quasiparticles at all times. It turns out that the minimal conductivity (at charge neutrality) is eventually determined by a marginally irrelevant operator, namely long-range Coulomb interaction. The critical transport to leading order in temperature turns out to be identical to that of a gas of hot Dirac electrons interacting solely via long-range Coulomb interaction~\cite{FSMS,Kashuba,MFS,MSF,Sheehy,FosterAleiner}.

(2) Within model II we find a universal temperature independent universal conductivity as has to be expected from dimensional reasoning. However, we find that the prefactor of the inverse relaxation time is extremely small compared to the one associated with the marginally irrelevant long-range Coulomb interaction. In contrast to model I, the scattering from the CDW (SDW) order parameter fluctuations is not completely inactive, but still extremely small. A simple order of magnitude estimate suggests that for all experimental purposes the universal quantum-critical conductivity is masked by the current relaxation due to long-range Coulomb interaction. 

We comment on the seeming discrepancy between the transport properties of model I and model II and argue that the two different pictures are actually compatible and describe the same basic physics, with model I overestimating the kinematical blocking.

On a more technical note another interesting result obtains within model I, in which bosons and fermions are directly coupled, namely the absence of boson-drag effects for the coupled Boltzmann equations. Boson-drag constitutes a serious complication in obtaining transport coefficients in field theories of electrons coupled to bosons, where the bosons are an effective degree of freedom, whose dynamics itself is governed by the underlying electrons. An often employed approximation for the calculation of electronic transport properties in such systems is that the bosonic system is assumed to be equilibrated on timescales relevant for electronic transport and drag effects are neglected. This approximation turns out to be exact in our case.

Finally, the crossover behavior of the conductivity for the the quantum-critical-to-insulator crossover is discussed for both forms of collective ordering, namely the CDW and SDW. As we argued before, in the case of CDW an Ising degree of freedom condenses, which is allowed at finite temperature. Thus the electronic degrees of freedom have a hard gap, which entails an exponentially suppressed conductivity. In the case of the SDW, on the other hand, a real quasiparticle gap only opens at temperature $T=0$ and finite temperature behavior is not governed by a hard gap, but rather by a "pseudogap", leading to a power-law suppression of the conductivity. 

Both scenarios allow to determine crossover curves which show remarkable similarities to experiments performed on suspended high-mobility graphene samples~\cite{Bolotin2}. It appears that they are qualitatively and quantitatively widely compatible with a recent analysis of the data sets under the hypothesis of an underlying electronic quasiparticle gap and might thus be a natural starting point for the identification of the fitting parameters discussed in Ref.~\onlinecite{Drut2}.

\subsection{Outline}
In Sec.~\ref{Sec:model} we introduce the scenario for the semimetal-to-insulator transition and motivate two different phenomenological models, called model I and model II. During the course of the paper we will switch between model I and model II, since different aspects of the problem are more conveniently discussed in one or the other formulation. 

We review the renormalization group treatment of model I following Ref.~\onlinecite{Herbut1} and also the large-N treatment of model II following Ref.~\onlinecite{Herbut3}. We furthermore stress some of the main physical properties of the models. In Sec.~\ref{Sec:Boltzmann} we introduce the general formalism of the Boltzmann approach. We apply the formalism to both models. In Sec.~\ref{Sec:BoltzI} we consider model I and solve a system of coupled Boltzmann equations for the fermionic and bosonic degrees of freedom. The bosonic degrees of freedom capture the physics of the collective instability and are eventually responsible for the opening of the electronic quasiparticle gap. We explicitly show that in this system of coupled transport equations neglecting the effect of "boson-drag" is {\it exact} to linear order in the applied electric field and proceed to analyze the Boltzmann equation for the massless Dirac particles. We find that inelastic scattering due to the Yukawa-coupling is kinematically forbidden (see Sec.~\ref{Sec:noleadingorder}) to {\it all} orders in perturbation theory. This results in the absence of a temperature independent minimal d.c. conductivity. In Sec.~\ref{Sec:BoltzII} we solve the Boltzmann equation of model II. In contrast to model I we find that there is a finite universal conductivity which does not depend upon temperature. However, we find that the prefactor of the inverse scattering time is tiny. This leads us to Sec.~\ref{Sec:Boltzcomp} where we compare the results of in model I with those in model II and argue in which sense the two results are compatible. In Sec.~\ref{Sec:QCRcon} we discuss the minimal conductivity in the quantum-critical regime and make an estimate of crossover temperatures. In a final section (Sec.\ref{Sec:discussion}) we discuss crossover behaviors which obtain upon lowering of temperature. Furthermore, we discuss the fundamental difference between the CDW and SDW cases with a special eye on pseudogap behavior in the SDW case. A discussion of experimental conditions and reference to puzzling results of recent experiments on suspended high mobility graphene samples concludes Sec.~\ref{Sec:discussion}.

\section{Spontaneous inversion symmetry breaking in the framework of two different models}\label{Sec:model}

It is well known that a tight binding description of non-interacting electrons on the honeycomb lattice in its low-energy version can be cast in the form of the Dirac theory for massless fermions. The Hamiltonian for the electronic part reads~\cite{NetoRMP}  
\begin{eqnarray}\label{Eq:Hamiltonian}
H=iv_F \int d^2x\bar{\Psi}_{\sigma,\kappa} \gamma_\mu \partial_\mu \Psi_{\sigma,\kappa}\;,
\end{eqnarray}
with $v_F$ being the Fermi velocity. We explicitly consider half-filling, {\it i.e.}, chemical potential $\mu=0$ throughout the paper. Double indices are henceforth summed over unless stated otherwise.
In the above notation we take $\Psi_{\sigma,\kappa}$ to be a spinor whose two components are associated with the two sublattices A and B. The index $\sigma$ denotes the spin degree of freedom whereas $\kappa$ denotes the valley ($K$ and $K'$ point in the Brillouin zone). The $\gamma$-matrices are defined as $\gamma_1= i\sigma_y$, and $\gamma_2= -i \sigma_x$, and $\bar{\Psi}^{\phantom{\dagger}}_{\sigma,\kappa}=\Psi^\dagger_{\sigma,\kappa} \sigma_z$, where $\sigma_{x,y,z}$ are the Pauli matrices acting in sublattice space.

Local interactions in this system can induce collective instabilities. Spontaneous symmetry breaking leads to a gap in the electronic single-particle excitation spectrum. We concentrate on order parameters which break the sublattice inversion symmetry, {\it i.e.}, the two sites per unit cell become chemically distinguishable. Two possible instabilities correspond to charge-density wave (CDW) and spin-density wave (SDW) states, whose order parameters in terms of electronic operators read
  \begin{equation}
  \chi=  (\chi_s, \vec{\chi}_t) = (\langle \bar{\Psi}_{\sigma,\kappa} \Psi_{\sigma,\kappa} \rangle,  \langle \bar{\Psi}_{\sigma,\kappa} \vec{\tau}_{\sigma \sigma'}\Psi_{\sigma',\kappa} \rangle )\;,
   \end{equation}
where $\vec{\tau}$ acts in spin space and is simply a vector composed of Pauli matrices. In the above representation the singlet $\chi_s$ corresponds to staggered charge density (CDW), and the triplet $\vec{\chi}_t$ to staggered local magnetization (SDW). Microscopically, a finite $\chi_s $  ($\vec{\chi}_t$) may be induced by a large nearest-neighbor (on-site) repulsion. In the following we explicitly study the case of CDW and comment on differences with the SDW case as we go along.

\subsection{Model I: Yukawa-theory and the description of the semimetal-to-insulator transition in the framework of an $\epsilon$-expansion}

 We start with introducing model I, which is a field theory of Dirac fermions coupled to a Landau-Ginzburg order parameter theory via a Yukawa-type coupling.
For the following discussions we generalize the model to $d$ dimensions and adopt a Lagrangian formulation. The generalization to $d$ dimensions eventually allows to treat the bosonic and fermionic sectors on an equal footing in a perturbative renormalization group treatment. We generalize the Dirac Hamiltonian to three dimensions such that matrix elements have a structure like in two dimensions: on a technical level this implies that we evaluate angular integrals as if they were in two dimensions, but keep the integrals over absolute values in $d$ dimensions. The following discussion closely mimics the main steps of Ref.~\onlinecite{Herbut1}.
We decompose the local interaction into separate dynamical fields via Hubbard-Stratonovich decoupling and formulate the effective action near the semimetal-to-insulator transition as $S=\int d\tau d\vec{x} L$, with $L=L_f + L_y + L_b+ L_c$, where the bosonic part is described by 
 \begin{eqnarray}
 L_b &=&    \frac{1}{2}   \chi_s \left (-\partial_\tau ^2 - v_s ^2 \nabla^2   + t \right) \chi_s  +   u_s (\chi_s \chi_s )^2  ,\nonumber \\
 \end{eqnarray}
and the Yukawa term that couples bosonic and fermionic fields reads
\begin{eqnarray}
 L_y =  g_s \chi_s \cdot \bar{\Psi}_{\sigma,\kappa}  \Psi_{\sigma,\kappa} \;.
\end{eqnarray}
We note that $L_b$ describes the CDW fluctuation mode, where $t$ is the tuning parameter, $u_s$ the self-interaction, and $v_s$ the velocity. It is thus a Landau-Ginzburg theory for an Ising order parameter. The physical meaning of the Yukawa term is that a CDW can decay into a pair of fermions.
We also introduce the inverse bosonic propagator at this point
\begin{eqnarray}\label{Eq:Bosonpropagator}
D^{-1}(x,y,\tau,\tau')=\left( -\partial_\tau^2-v_s^2 \nabla^2+t \right)\delta(x-y)\delta(\tau-\tau')  \;,\nonumber \\
\end{eqnarray}
since it is needed for the generic transport equations.
It is important to note here that $L$ lacks the Lorentz symmetry due to the bosonic velocity $v_{s}\neq v_F$.

Interestingly, the above Gross-Neveu theory is renormalizable in $3+1$ dimensions, where {\it both} the Yukawa coupling $g_s$ and the self-interaction coupling $u_s$ become dimensionless. It is then possible to study the effective Yukawa theory in $d=3-\epsilon$ dimensions using renormalization group (RG) methods with $\epsilon$ being the control parameter for the perturbative series ($\epsilon=1$ corresponds to the two-dimensional model). We note that other authors have considered similar field-theories using an RG framework in the context of d-wave superconductors~\cite{VojtaSachdev}.

\subsubsection{RG description of the semimetal-to-insulator transition}

In the following discussion we sketch the RG treatment of Ref.~\onlinecite{Herbut1}. We first restore Lorentz invariance of the model, implying we set $v_s=v_F=1$ throughout this paragraph. This is a crucial point and we extensively comment on this step below. 

\begin{figure}
\includegraphics[width=0.45\textwidth]{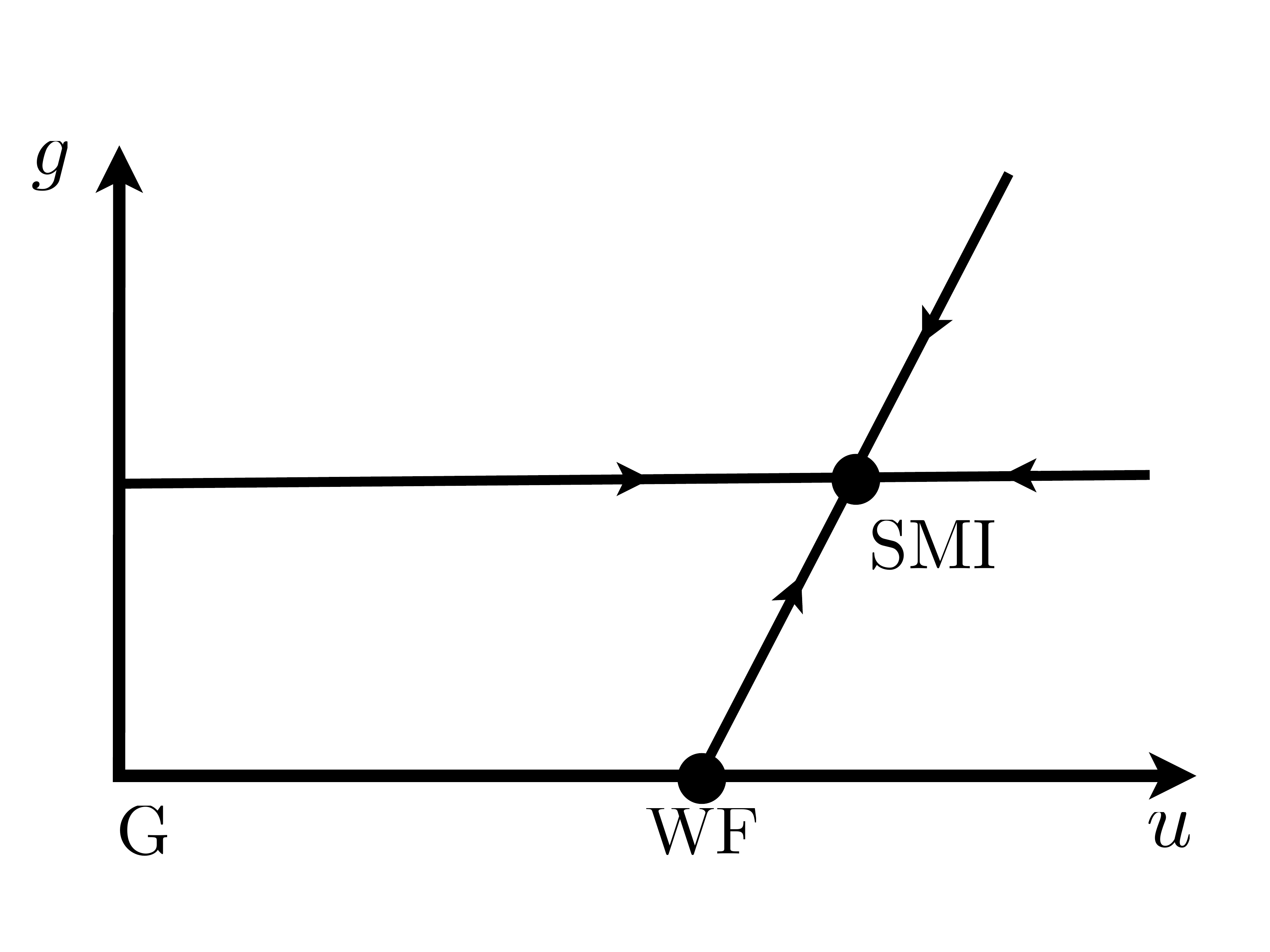}
\caption{Schematic RG flow diagram: $g$ and $u$ are the Yukawa and the bosonic coupling, respectively; G denotes the unstable Gaussian fixed point of the Landau theory and WF the Wilson-Fisher fixed point, which is unstable towards SMI (semimetal-to-insulator fixed point) upon switching on the Yukawa coupling $g$. The transition is tuned by the bosonic mass parameter $t$ which is perpendicular to the $u-g$ plane and not shown explicitly.}\label{Fig:flow}
\end{figure}

We take $L$ as the starting point and integrate out fermionic and bosonic modes within the momentum-shell defined by $\Lambda/b<(\omega^2+\vk^2)^{1/2}<\Lambda$. We furthermore introduce dimensionless couplings $g=\frac{g_s}{8\pi^2 \Lambda^\epsilon}$ and $u=\frac{u_s}{8\pi^2\Lambda^{\epsilon}}$ with $\epsilon=3-d$ yielding
\begin{eqnarray}
\frac{dg^2}{d \ln b}&=&g^2(\epsilon-7g^2)\quad \rm{and} \nonumber \\ \frac{d u}{d\ln b}&=&u(\epsilon-8g^2)-36 u^2+2g^4\;,
\end{eqnarray}
to one-loop order. The above flow equations obtain for the bosonic mass $t$ tuned to criticality. 
The intricate flow diagram is schematically depicted in Fig.~\ref{Fig:flow}.
Most interestingly, the usual Wilson-Fisher fixed point (WF) of the $\phi^4$-theory is unstable towards finite Yukawa coupling $g$ leading to an additional non-trivial fixed point called SMI located at
\begin{eqnarray}\label{Eq:fixedpoint}
g^*=\sqrt{\frac{\epsilon}{7}}\quad {\rm{and}}\quad u^*=\frac{2}{63}\epsilon\;.
\end{eqnarray}
This fixed point describes the semimetal-to-insulator transition and the associated tuning parameter is the bosonic mass parameter $t$ (Eq.~\eqref{Eq:Bosonpropagator}). The correct theory thus is described by electrons and bosons. A simple order parameter theory would not suffice to describe the physics correctly, since there the critical point would essentially be given by WF. 

In the above picture we have assumed the equivalence of bosonic and fermionic velocities, {\it i.e.}, Lorentz invariance. A closer look at the theory reveals that this is only asymptotically correct and breaking the relativistic invariance actually corresponds to an irrelevant perturbation with respect to the fixed point SMI. The corresponding RG equation was derived in Ref.~\onlinecite{Herbut1} and reads
\begin{eqnarray}\label{Eq:flowvelocities}
\frac{d \delta}{d \ln b}=-\frac{4\epsilon}{7}\delta
\end{eqnarray}
at $g^*$, where $\delta=\frac{v_F-v_b}{v_F}$. This parameter will play a vital role in the analysis of the leading scattering mechanism in Sec.~\ref{Sec:noleadingorder}.

We note that in the case of SDW the RG equations remain structurally intact, with modified combinatorial factors coming from the order parameter being $O(3)$ instead of $Z_2$.
\subsubsection{Long-range tail of Coulomb interaction}\label{Sec:LRC}

It is know that in a half-filled electronic system on the honeycomb lattice long-range Coulomb interaction is marginally irrelevant~\cite{Guinea}. It is thus also important to understand the role of long-range Coulomb interaction at SMI. In order to do this one has to generalize the Coulomb interaction to $d$ dimension following Ref.~\onlinecite{Herbut1,Herbut2} 
\begin{eqnarray}\label{Eq:CoulombInt}
L_c=ia_0\bar{\Psi}_{\sigma,\kappa}\sigma_z \Psi_{\sigma,\kappa}+\frac{1}{2e^2}a_0|\nabla|^{d-1}a_0 \;. 
\end{eqnarray} 
This form ensures that integration over $a_0$ induces a $\propto \frac{1}{r}$ density-density interaction between fermions in any dimension. It is found~\cite{Herbut1} that Coulomb interaction obeys the following RG equation
\begin{eqnarray}\label{Eq:RGCoulomb}
\frac{de^2}{d \ln b}=-\frac{32 \pi^2}{3}(2\delta_{d,3}+1)e^4\;.
\end{eqnarray}
One observes that our choice of $L_c$ (Eq.~\eqref{Eq:CoulombInt}) indeed produces a marginally irrelevant long-range interaction in any dimension. This establishes long-range Coulomb interaction as the least irrelevant coupling at SMI.

\subsubsection{Physics of SMI}

The physics of the quantum-critical point SMI has been analyzed in great detail by Herbut {\it et al.}~\cite{Herbut1} and we only repeat one key feature. A prediction of the field theory is that right at the quantum-critical point the fermion propagator behaves as 
\begin{eqnarray}
G_f^{-1}\propto (\omega^2+k^2)^{\frac{1-\eta_f}{2}}
\end{eqnarray}
where $\eta_f=\frac{3\epsilon}{14}$ is the anomalous dimension. In this sense there are no well-defined fermionic quasiparticles at the critical point. However, it was shown by Damle and Sachdev~\cite{Damle} that a Boltzmann transport analysis in terms of quasiparticles still captures the relevant physics in the hydrodynamic regime ($\hbar \omega \ll k_B T$). In that sense the semiclassical Boltzmann equation only constitutes a convenient starting point to solve the more complicated Quantum-Boltzmann equation, which we refrain from showing in full bloom in this paper.

\subsection{Model II: the Gross-Neveu model and description of the semimetal-to-insulator transition in large-N}

In the following we consider an alternative formulation of the physics described above. The advantage of the below model is that we can analyze the model directly in two dimensions. In order to do so we employ a generalization in the spirit of the large-N method. The large-N limit is obtained by generalizing the fermionic spinors with a new index $i=1,...,N$. The model of the free electrons was introduced before in Eq.~\eqref{Eq:Hamiltonian}. In the formulation of the model we follow Ref.~\onlinecite{Herbut3}. The interaction is now modeled by contact interaction of the type
\begin{eqnarray}\label{Eq:Lint}
L_{\rm{int}} &=& \frac{g_s}{N} \left( \bar{\Psi}_{\sigma,\kappa,i} \Psi_{\sigma,\kappa,i}\right)^2 \nonumber \\ &+& \frac{g_t}{N} \left( \bar{\Psi}_{\sigma,\kappa,i} \tau_{\sigma\sigma'}\Psi_{\sigma',\kappa,i}\right)^2 \;.
\end{eqnarray}
The physical equivalence of the above model with model I is established upon noting that $g_s$ drives an instability towards CDW, whereas $g_t$ drives the instability towards SDW.
\subsubsection{Quantum-criticality on the level of large-N}

In the following we again concentrate on the case of the CDW, implying that we set $g_t=0$. For large $N$, $g_s$ can give rise to a dynamically generated mass of the type 
\begin{eqnarray}
m\propto \langle  \bar{\Psi}_{\sigma,\kappa,i} \Psi_{\sigma,\kappa,i} \rangle \;,
\end{eqnarray}
which is determined by the gap equation
\begin{eqnarray}
-\frac{1}{g_s}=8\int_{-\infty}^{\infty} d \omega\int^{\Lambda/v_{F}}  \frac{d^2k}{(2\pi)^3}\frac{1}{\omega^2+v_F^2 k^2+m^2}
\end{eqnarray}
which can be integrated to yield
\begin{eqnarray}\label{Eq:largeN}
1=\frac{2g_s\Lambda}{v_F^2 \pi}\left( \sqrt{\frac{m^2}{\Lambda^2}} -\sqrt{\frac{m^2+\Lambda^2}{\Lambda^2}} \right)\;,
\end{eqnarray}
where $\Lambda \gg m$ is the ultraviolet cutoff. Demanding invariance of $m$ under rescaling $\Lambda \to \frac{\Lambda}{b}$ we obtain
\begin{eqnarray}
\frac{dg}{d \ln b}=-g-g^2
\end{eqnarray}
where $\frac{2g_s\Lambda}{v_F^2 \pi}=g$ was used (note that Eq.~\eqref{Eq:largeN} differs from Ref.~\onlinecite{Herbut3} due to a slightly different choice in the cutoff scheme). $g^\star=-1$ represents a quantum-critical point, where for $g<-1$ runaway flow into a phase with broken chiral symmetry, {\it i.e.}, $m\neq0$ obtains. This model represents an alternative starting point for studying the critical transport properties in the framework of the Boltzmann equation.

\subsubsection{The role of long-range Coulomb interaction}

Without going into the details of the derivation~\cite{Herbut3} we simply state that the effect of long-range Coulomb interaction has also been studied in the context of this model in the limit of large-N. In order to do so the Coulomb interaction~\eqref{Eq:CoulombInt} for $d=2$ is equally generalized to large $N$ and the flow equation of the charge reads
\begin{eqnarray}
\frac{de^2}{d\ln b}=-\frac{e^4}{2\pi N}
\end{eqnarray}
which confirms that the long-range Coulomb interaction is marginally irrelevant, in qualitative agreement with Eq.~\eqref{Eq:RGCoulomb} from model I and Ref.~\onlinecite{Guinea}.

\vspace{5mm}
We close this section with a generic comment about recent numerical simulations performed for Dirac fermions with only long-range interactions in terms of a lattice gauge theory simulation. It was shown~\cite{Drut} that even without short-range interactions a quantum phase transitions as a function of the fine-structure constant can be driven from a semimetal to an insulator, where in the insulating phase chiral symmetry is observed to be broken. This transition happens at a critical value of the fine structure constant and numerical values for this critical coupling come close to the value expected in vacuum. Within numerical accuracy the transition proves to be second order and more remarkably the critical exponents seem compatible with the above two models with short-range interactions exclusively. 

We note that for a purely Hubbard type model there has been a recent work discussing the Mott transition which is not driven by collective symmetry breaking, but a transition into an intermediate paramagnetic spin liquid phase is observed~\cite{Zheng}. Only upon increasing the local interaction the model has a transition into an antiferromagnetic phase. The nature of this spin liquid phase is currently not fully understood and not subject of our discussion.

\section{Boltzmann transport equation}\label{Sec:Boltzmann}

The Boltzmann transport equation in combination with RG methods and large-N has proven a valuable tool in analyzing transport properties of quantum-critical systems~\cite{ssbook,Damle,ssqhe,bhaseen,FSMS}.
In this section we introduce the basic notation and concepts for the calculation of the critical conductivity following closely Ref.~\onlinecite{FSMS}.

It is well known that in momentum space one can diagonalize the Hamiltonian~\eqref{Eq:Hamiltonian} using a unitary matrix of the form
\begin{eqnarray}
\hat{U}^{\phantom{-1}}_{\vk}=\frac{1}{\sqrt{2}k} \left( \begin{array}{cc}  K^* & -K^* \\ k& k\end{array} \right)
\end{eqnarray}
with
\begin{eqnarray}
K\equiv k_{x}+ik_{y}
\end{eqnarray}
and $k=|\mathbf{k}|=|K|$. Expressing the Hamiltonian $H_{0}$ in terms of particle and hole operators
\begin{eqnarray}
\left( \begin{array}{c} \gamma_{+,\sigma,\kappa}(\vk)  \\ \gamma_{-,\sigma,\kappa}(\vk)  \end{array} \right) = \hat{U}^{-1}_{\vk}\Psi_{\sigma,\kappa} (\vk) 
\end{eqnarray}
we obtain the diagonal form
\begin{equation}
H_{0}=\sum_{\lambda=\pm}\int \frac{d^{2}k}{(2\pi )^{2}}\lambda k\, \gamma
_{\lambda,\sigma,\kappa }^{\dagger }(\mathbf{k})\gamma^{\phantom{\dagger}} _{\lambda,\sigma,\kappa}(\mathbf{k}),
\end{equation}%
where the sum $\lambda $ extends over electron- and hole-band, denoted $+,-$.

We can also express the electrical current in terms
of the $\gamma _{\pm }$. For the case of a spatially independent current the result can be written as
\begin{eqnarray}
\mathbf{J}=\mathbf{J}_{I}+\mathbf{J}_{II} 
\end{eqnarray}%
with
\begin{equation}
\mathbf{J}_{I}=e\sum_{\lambda=\pm}\int \frac{d^{2}k}{(2\pi )^{2}}\frac{%
\lambda \mathbf{k}}{k}\gamma _{\lambda,\sigma,\kappa}^{\dagger }(\mathbf{k})\gamma^{\phantom{\dagger}}
_{\lambda,\sigma,\kappa}(\mathbf{k})\;. \label{defj1}
\end{equation}%
$\mathbf{J}_{I}$ measures the current carried by the motion of the
quasiparticles and quasiholes; the prefactor $\lambda $ accounts for these excitations having opposite charges. The operator $\mathbf{J}_{II}$ is not shown explicitly; it describes the
creation of a quasiparticle-quasihole pair (it corresponds to the so-called Zitterbewegung, see Ref.~\onlinecite{NetoRMP}) 

As in the problems studied in Refs.~\onlinecite{ssbook,Damle,ssqhe,bhaseen,FSMS}, in a particle-hole symmetric situation a current carrying state with
holes and electrons moving in opposite directions has a
vanishing total momentum and the current can decay by creation or annihilation of particle-hole pairs without violating momentum conservation. This is the physical reason why at the
particle-hole symmetric point, {\it i.e.}, at vanishing  deviation of the chemical
potential from the Dirac point, the d.c. conductivity is finite even in the
absence of momentum relaxing impurities. However, at finite
deviation from particle-hole symmetry, driven by chemical potential or the presence of second neighbor hopping $t'$, a driving electric field always excites
the system into a state with finite momentum which cannot decay. This 
entails an infinite d.c. conductivity in absence of Umklapp scattering or disorder. 

As a first step one defines the distribution function
\begin{eqnarray}
f_{\lambda }(\mathbf{k},t)=\left\langle \gamma _{\lambda,\sigma,\kappa}^{\dagger }(%
\mathbf{k},t)\gamma^{\phantom{\dagger}}_{\lambda,\sigma,\kappa}(\mathbf{k},t)\right\rangle .  \label{defg}
\end{eqnarray}
Note, there is no sum over $\sigma,\kappa$ on the r.h.s.. We furthermore assume that the distribution
functions are the same for all valleys and spins, which has to be expected for symmetry reasons. In equilibrium, {\it i.e.}, in the
absence of external perturbations, the distribution functions are Fermi functions
\begin{eqnarray}
f_{+}(\mathbf{k},t) &=&f^{0}(k)= \frac{1}{e^{( k -\mu)/T}+1}\;,
\notag \\
f_{-}(\mathbf{k},t) &=&f^{0}(-k) =
\frac{1}{e^{(-k-\mu)/T}+1}\;,
\end{eqnarray}%
where we temporarily allow for a finite chemical potential $\mu$.

In principle, off-diagonal elements such as $\langle \gamma^\dagger_{\pm} \gamma^{\phantom{\dagger}}_{\mp}\rangle$ are also created by an electric field. However, they are not needed to evaluate ${\bf{J}}_{I}$, which is the part of the current that we focus on in the hydrodynamic regime, $\omega\ll k_B T$. Furthermore, the off-diagonal elements feed back to the kinetic equation of the diagonal elements only to higher order in perturbation theory.
In the presence of an external electric field $\mathbf{E%
}$ acting as a driving force on quasielectrons and quasiholes, we find the semiclassical Boltzmann equation
\begin{equation}
\left( \frac{\partial }{\partial t}+e\mathbf{E}\cdot \frac{\partial }{%
\partial \mathbf{k}}\right) f_{\lambda }(\mathbf{k},t)=I_{\rm coll}\;,  \label{trans0}
\end{equation}
where $I_{\rm coll}$ is the scattering integral. We parameterize the change in $f_{\lambda }$ from its equilibrium value by \cite{Ziman}
\begin{eqnarray}
f_{\lambda }(\mathbf{k},\omega )&=&2\pi \delta (\omega )f^{0}(\lambda
k) \nonumber \\ &+& \lambda ev_F \frac{\mathbf{k}\cdot \mathbf{E}(\omega )}{k}f^{0}(\lambda
k)(1-f^{0}(\lambda k))g(k,\omega )\;,\nonumber \\  \label{paramet}
\end{eqnarray}%
where we have performed a Fourier transform in time to frequencies, $\omega $%
, and introduced the unknown function $g(k,\omega )$.
At the particle-hole symmetric point ($\mu=0$), an applied electric field
generates deviations in the distribution functions having opposite sign for
quasiparticles and quasiholes. Formally, this is a consequence of the driving term in Eq.~\eqref{trans0} being asymmetric under
$\lambda \to -\lambda$, and thus the solution has to be asymmetric as well.
This reflects the fact that there is an increased number of quasiholes and
quasiparticles  moving parallel and antiparallel to the exciting field, respectively. As
quasiparticles and -holes have opposite charges, their electrical currents are
equal and add up, while their net momenta have opposing signs and subtract to yield zero. In this paper we consider instabilities, which do not break particle-hole symmetry, and thus the above parametrization is strictly justified.

\subsection{Boltzmann equations for model I}\label{Sec:BoltzI}

We can formulate the Boltzmann equation for the electrons and holes (we only show the equation for the electrons; for holes it follows by symmetry) to lowest order in the electron-boson coupling as
\begin{widetext}
\begin{eqnarray}\label{Eq:BoltzDirac}
e \frac{\bf{k}}{k}\cdot{\bf{E}} \partial_k f^0(k)&=& 4 g_s^2\int \frac{d^d q}{(2\pi)^d}M_{++}(\vk,\vk-{\bf{q}})\;{\rm{Im}}\;D({\bf{q}},k-|\vk-{\bf{q}}|)\times\nonumber \\ &\times&\left( n_b(k-|\vk-{\bf{q}}|)\left(f_+(\vk)-f_+(\vkq)\right) +f_+(\vk)\left( 1-f_+(\vkq)\right)\right) \nonumber \\ &+&4 g_s^2\int \frac{d^d q}{(2\pi)^d}M_{+-}(\vk,\vk-{\bf{q}})\;{\rm{Im}}\;D({\bf{q}},k+|\vk-{\bf{q}}|)\times\nonumber \\ &\times&\left( n_b(k+|\vk-{\bf{q}}|)\left(f_+(\vk)-f_-(\vkq)\right) +f_+(\vk)\left( 1-f_-(\vkq)\right)\right)\;,
\end{eqnarray}
\end{widetext}
where
\begin{eqnarray}
M^{\phantom{-1}}_{\lambda \lambda'}(\vk,\vk')=T^{\phantom{-1}}_{\lambda \lambda'}(\vk,\vk')T^{-1}_{\lambda'\lambda}(\vk,\vk')
\end{eqnarray}
with
\begin{eqnarray}\label{Eq:matrixelement}
T^{\phantom{-1}}_{\lambda \lambda'}(\vk,\vk')&=&\left(\hat{U}^{{-1}}_{\vk} \tau_z \hat{U}^{\phantom{-}}_{\vk'}\right)_{\lambda \lambda'} \nonumber \\ &=& -\frac{1}{2}\left(1-\lambda \lambda' \frac{K\left( K'\right)^*}{k k'} \right)\;.
\end{eqnarray}
$D$ is the bosonic propagator~\eqref{Eq:Bosonpropagator} and $n_b(\epsilon)$ the Bose distribution function.

There exists an analogous equation for the bosonic degrees of freedom. Since the bosons do not couple to the electromagnetic field, the driving term (l.h.s.) is exactly zero. However, the scattering integral couples the non-equilibrium fermions to the bosons constraining the bosonic distribution function to solve
\begin{widetext}
\begin{eqnarray}\label{Eq:Boltzbosons}
0&=&\int \frac{d^d q}{(2\pi)^d}M_{++}(\vk,\vk+{\bf{q}})\left(n_b(\vk)(f_+(\vk+{\bf{q}})+f_-(\vk+{\bf{q}})-\left(f_+({\bf{q}})+f_-({\bf{q}}) \right)   ) \right)\nonumber \\ &+&\int \frac{d^d q}{(2\pi)^d}M_{+-}(\vk,\vk+{\bf{q}})\left(n_b(\vk)(f_+(\vk+{\bf{q}})+f_-(\vk+{\bf{q}})-\left(f_+({\bf{q}})+f_-({\bf{q}}) \right)   ) \right)\nonumber \\ &+&\int \frac{d^d q}{(2\pi)^d}M_{++}(\vk,\vk+{\bf{q}})\left( f_+({\bf{k}}+{\bf{q}})(1-f_+({\bf{q}}))+f_-({\bf{k}}+{\bf{q}})(1-f_-({\bf{q}})) \right)\nonumber \\&+&\int \frac{d^d q}{(2\pi)^d}M_{+-}(\vk,\vk+{\bf{q}})\left( f_+({\bf{k}}+{\bf{q}})(1-f_-({\bf{q}}))+f_-({\bf{k}}+{\bf{q}})(1-f_+({\bf{q}})) \right)\;.
\end{eqnarray}
\end{widetext}
This bosonic Boltzmann equation~\eqref{Eq:Boltzbosons} has to be solved simultaneously with the fermionic one~\eqref{Eq:BoltzDirac}. In principle, there is also a boson only scattering term coming from the self-interaction ($\chi_s^4$-term)~\cite{Damle}. This term is not shown explicitly for reasons of conciseness. It turns out that its neglect proves to be exact in the case at hand. This is rooted in the fact that the bosons remain in equilibrium, which is shown explicitly below.

\subsubsection{The effect of 'boson-drag'}

Severe complications in calculating electromagnetic response functions in systems in which fermions and collective (slave-)bosons are coupled to each other result from the fact that one has to solve coupled Boltzmann equations (Eq.~\eqref{Eq:BoltzDirac} and Eq.~\eqref{Eq:Boltzbosons}). The physical reason for that is simple: electrons transfer momentum to the bosons, which themselves are driven out of equilibrium. Since the bosons also scatter from the fermions, there is a feedback effect which in principle has to be taken seriously. Often this is circumvented by assuming the bosonic sector to be in equilibrium, which is equivalent to saying that the bosonic sector equilibrates due to lattice effects or impurities on a time scale faster than the electronic scattering time. This can be an oversimplifying assumption with no {\it a priori} justification. One can appreciate this most easily by considering a Galilean invariant electronic system interacting with effective bosonic modes.  In this situation neglecting the generic non-equilibrium situation for the bosonic sector yields fundamentally wrong results, namely a {\it finite} d.c. conductivity. One can convince oneself that taking into account the Boltzmann equation for the bosons reestablishes the physically correct result of infinite d.c. conductivity (see chapter on 'phonon-drag' in Ref.~\onlinecite{Ziman}). 

We now turn our attention to Eq.~\eqref{Eq:Boltzbosons} and expand the fermionic and bosonic distribution functions to linear order in the applied external field ${\bf{E}}$ using Eq.~\eqref{paramet}. For the bosonic sector we stick to the symbolic
\begin{eqnarray}
n_b(\vk)=n_b^0(\epsilon(\vk))+u_{\vk}\Phi(k)
\end{eqnarray}
with $u_{\vk}$ being linear in the applied field and $n_b^0(\epsilon)$ as the equilibrium Bose distribution function.
A simple analysis shows that $\Phi(k)$ drops out of Eq.~\eqref{Eq:Boltzbosons} and the remaining parts annihilate exactly, as they should for consistency.
The above scattering integral is thus solved exactly by letting
\begin{eqnarray}
n_b(\vk)=n_b^0(\epsilon(\vk))  
\end{eqnarray}
 and using Eq.~\eqref{paramet} for the fermions. We have thus shown that at the Dirac point the 'boson-drag' vanishes exactly and considering the bosons in thermodynamic equilibrium is exact.

One can convince oneself that in a situation with finite chemical potential $\mu$ this ceases to be true and the bosons have to be treated as being out of equilibrium. In such a situation a finite amount of disorder is needed to relax the net momentum and the associated current.  

 We have thus established that in clean graphene at $\mu=0$ the bosonic sector remains in equilibrium. We can understand this more intuitively as follows: we first consider an electronic band with quadratic dispersion interacting with phonons. In this situation the bosonic sector absorbs some momentum from the electrons which scatter and is itself driven out of equilibrium. In graphene we have no quadratic disperion and by virtue of the perfect particle-hole symmetry ($\mu=0$) the charge carrying modes at the Dirac point are modes with net momentum zero. Thus the bosons do not absorb any momentum which they have to dissipate. This implies they are not driven out of equilibrium and treating the bosons in equilibrium is no approximation but exact to linear order in the applied external field ${\bf {E}}$. Note that the above reasoning is exclusive to electrical transport and does not hold for thermal transport.

This line of argument is not exclusive to graphene or the matrix elements of the above interaction, but also applies to other systems with perfect particle-hole symmetry which is unbroken by the interaction mediated by the bosons. As a timely example one could consider bilayer graphene~\cite{NetoRMP} where the quasiparticles interact with effective bosonic degrees of freedom and the reasoning still holds.

\subsubsection{Cancellation to infinite order}\label{Sec:noleadingorder}

\begin{figure}
\includegraphics[width=0.35\textwidth]{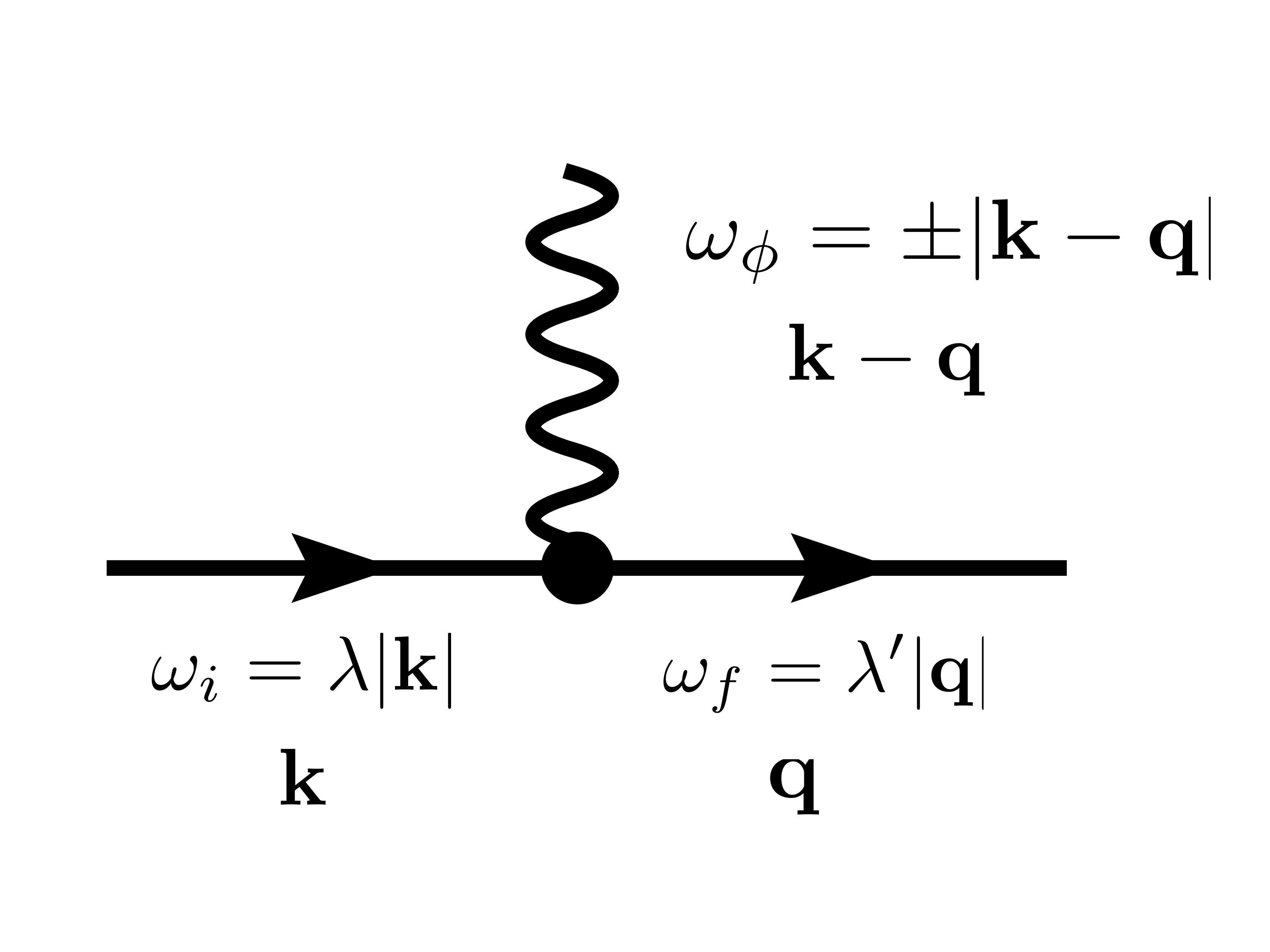}
\caption{Momentum and energy structure of the Yukawa vertex $g$ for scattering of quasiholes ($\lambda=-$, full line) and quasielectrons ($\lambda=-$) from the collective mode (wiggly line). One can work out that for $v_s=v_F$ the bosonic spectral function is composed of sharp quasiparticles and only collinear scattering is allowed; collinear scattering, however, annihilates the associated matrix elements.}\label{Fig:vertex}
\end{figure}

For dimensional reasons one expects the solution to Eq.~\eqref{Eq:BoltzDirac} to yield a quantum-critical conductivity which is independent of temperature~\cite{Damle,ssqhe,bhaseen}. It turns out that this is not true. To quadratic order in the interaction $g$ the scattering integral~\eqref{Eq:BoltzDirac} is exactly zero due to a conspiracy of the matrix element and energy conservation. This in itself does not rule out a temperature independent conductivity, since contributions to all orders in $g$ provide scattering times independent of temperature.  However, one can convince oneself that the kinematical constraint shown in Fig.~\ref{Fig:vertex} annihilates contributions to current relaxation due to the Yukawa coupling $g$ to {\it all} orders in perturbation theory, provided $v_F=v_s$. This can easily be seen by looking at the matrix elements Eq.~\eqref{Eq:matrixelement} and imposing energy and momentum conservation as shown in Fig.~\ref{Fig:vertex}. The exact cancellation is only operational for $v_s=v_F$. Breaking of Lorentz invariance, {\it i.e.}, allowing for $v_s\neq v_F$ allows to relax the current. This implies that current relaxation due to the collective CDW degree of freedom is determined not only by $g_s$ but also by an irrelevant parameter, namely $\delta$, which was introduced as the dimensionless parameter measuring the breaking of Lorentz invariance. Since $g_s$ assumes its fixed point value, which is a pure number, the inverse scattering time vanishes upon lowering temperature at least like $\delta \propto T^{4/7}$, see Eq.~\eqref{Eq:flowvelocities}, leading to a diverging d.c. conductivity. One now has to determine the least irrelevant scattering mechanism among all possible ones. As we argued before, all scattering events coming from the interaction with the collective mode scale with additional powers of $\delta$ and thus are irrelevant. 

We have seen in Sec.~\ref{Sec:LRC} that long-range Coulomb interaction constitutes another perturbation to SMI, which is only marginally irrelevant and consequently vanishes more slowly as temperature is lowered, namely in a logarithmic manner. From the above analysis we conclude that the dominant current relaxation mechanism stems from long-range Coulomb interaction. This implies that to leading order the behavior in the semimetal is the same as at SMI. 

This statement in itself is very surprising. We are now going to reexamine and reinterpret it in the framework of model II.

\subsection{Boltzmann equation for model II}\label{Sec:BoltzII}

The following analysis is carried out directly in 2 dimensions starting from the interaction term Eq.~\eqref{Eq:Lint}. The collision term on the r.h.s of Eq.~\eqref{trans0} can be determined using Fermi's golden rule or equivalently from a Quantum-Boltzmann equation. We refrain from an explicit derivation and refer the reader to Ref.~\cite{FSMS}, which gives an explicit derivation for the case of long-range interactions which is easily generalized to the present case. The generalization leads to

\begin{widetext}
\begin{eqnarray}
&&\left( \frac{\partial }{\partial t}+e\mathbf{E}\cdot \frac{\partial }{%
\partial \mathbf{k}}\right) f_{\lambda }(\mathbf{k},t)=-\frac{(2\pi )}{v_{F}}%
\int \frac{d^{2}k_{1}}{(2\pi )^{2}}\frac{d^{2}q}{(2\pi )^{2}}\Biggl\{  \notag
\\
&&\delta (k-k_{1}-|\mathbf{k}+\mathbf{q}|+|\mathbf{k}_{1}-\mathbf{q}|)R_{1}(%
\mathbf{k},\mathbf{k}_{1},\mathbf{q})\Bigl\{f_{\lambda }(\mathbf{k}%
,t)f_{-\lambda }(\mathbf{k}_{1},t)[1-f_{\lambda }(\mathbf{k}+\mathbf{q},t)]
\notag \\
&&~~~~~~\times \lbrack 1-f_{-\lambda }(\mathbf{k}_{1}-\mathbf{q}%
,t)]-[1-f_{\lambda }(\mathbf{k},t)][1-f_{-\lambda }(\mathbf{k}%
_{1},t)]f_{\lambda }(\mathbf{k}+\mathbf{q},t)f_{-\lambda }(\mathbf{k}_{1}-%
\mathbf{q},t)\Bigr\}  \notag \\
&&\delta (k+k_{1}-|\mathbf{k}+\mathbf{q}|-|\mathbf{k}_{1}-\mathbf{q}|)R_{2}(%
\mathbf{k},\mathbf{k}_{1},\mathbf{q})\Bigl\{f_{\lambda }(\mathbf{k}%
,t)f_{\lambda }(\mathbf{k}_{1},t)[1-f_{\lambda }(\mathbf{k}+\mathbf{q},t)] \\
&&~~~~~~\times \lbrack 1-f_{\lambda }(\mathbf{k}_{1}-\mathbf{q}%
,t)]-[1-f_{\lambda }(\mathbf{k},t)][1-f_{\lambda }(\mathbf{k}%
_{1},t)]f_{\lambda }(\mathbf{k}+\mathbf{q},t)f_{\lambda }(\mathbf{k}_{1}-%
\mathbf{q},t)\Bigr\}\Biggr\},  \notag  \label{trans1}
\end{eqnarray}%
where
\begin{eqnarray}
R_{1}(\mathbf{k},\mathbf{k}_{1},\mathbf{q}) &=&4N\left( \bigl|\mathcal{T}_{+--+}(\mathbf{k},\mathbf{k}_{1},\mathbf{q})\bigr|%
^{2}+\bigl|\mathcal{T}_{+-+-}(\mathbf{k},\mathbf{k}_{1},-\mathbf{k}-\mathbf{q}+%
\mathbf{k}_{1})\bigr|^{2}\right)\,,  \notag \\
R_{2}(\mathbf{k},\mathbf{k}_{1},\mathbf{q}) &=&4N \bigl|\mathcal{T}_{++++}(\mathbf{k},\mathbf{k}_{1},\mathbf{q})
\bigr|^{2}\;.  \label{deft12}
\end{eqnarray}
\end{widetext}
Note that the collision kernels $R_1$ and $R_2$ both come with a prefactor $N$. There are additional contributions coming from crossed diagrams, which are however down by a factor $1/N$, see Ref.~\onlinecite{FSMS}.
The terms proportional to $R_{1}$ represent collisions between oppositely
charged particles, while those proportional to $R_{2}$ are collisions
between like charges. Other processes, where a particle-hole pair is
created, turn out to have vanishing phase space: it is not possible to fulfill momentum and energy conservation at the same time. This is a peculiarity of the linear dispersion, {\it i.e.}, $\varepsilon_k=v_F k$ (see also Ref.~\onlinecite{FSMS}). The scattering matrix elements are given by
\begin{eqnarray}
\mathcal{T}_{\lambda\lambda_1\lambda_2\lambda_3}(\mathbf{k},\mathbf{k}_1,\mathbf{q})=\frac{g_s^2}{N^2} T_{\lambda \lambda_3}(\mathbf{k}+\mathbf{q},\mathbf{k})T_{\lambda_1 \lambda_2}(\mathbf{k}_1-\mathbf{q},\mathbf{k}_1)\;.\nonumber \\
\end{eqnarray}

We now proceed to the linearization of (\ref{trans1}) by inserting the
parametrization (\ref{paramet}) (we define $g_\lambda(k,\omega) \equiv \lambda g(k,\omega)$) and find
\begin{widetext}
\begin{eqnarray}
&& \frac{(-i \omega g_\lambda(k,\omega) - \lambda v_F/T)}{(e^{v_F k/T} + 1)
(e^{-v_F k/T} +
1)} \frac{\mathbf{k}}{k}= - \frac{(2 \pi)}{v_F} \int \frac{d^2 k_1}{(2 \pi)^2%
} \frac{d^2 q}{(2 \pi)^2} \Biggl\{  \notag \\
&&~~ \frac{\delta(k -k_1 - |\mathbf{k} + \mathbf{q}| + |\mathbf{k}_1 -
\mathbf{q}|) R_1 (\mathbf{k}, \mathbf{k}_1, \mathbf{q})}{(e^{-v_F k/T} +
1)(e^{v_F k_1/T}+1) (e^{v_F |\mathbf{k} + \mathbf{q}|/T} + 1) (e^{-v_F |%
\mathbf{k}_1 - \mathbf{q}|/T} + 1)}  \notag \\
&& \times \left[ \frac{\mathbf{k}}{k} g_\lambda(k, \omega) +
\frac{\mathbf{k}_1}{k_1}
g_{-\lambda} (k_1, \omega) - \frac{ (\mathbf{k} + \mathbf{q})}{|\mathbf{k} +
\mathbf{q}|%
} g_\lambda (|\mathbf{k} + \mathbf{q}|, \omega) - \frac{ (\mathbf{k}_1 -
\mathbf{q})%
}{|\mathbf{k}_1 - \mathbf{q}|} g_{-\lambda} (|\mathbf{k}_1 - \mathbf{q}|,
\omega) %
\right]  \notag \\
&&+ \frac{\delta(k + k_1 - |\mathbf{k} + \mathbf{q}| - |\mathbf{k}_1 -
\mathbf{q}|) R_2 (\mathbf{k}, \mathbf{k}_1, \mathbf{q})}{(e^{-v_F k/T} +
1)(e^{-v_F k_1/T}+1) (e^{v_F |\mathbf{k} + \mathbf{q}|/T} + 1) (e^{v_F |%
\mathbf{k}_1 - \mathbf{q}|/T} + 1)} \label{trans2} \\
&& \times \left[ \frac{\mathbf{k}}{k} g_\lambda(k, \omega) +
\frac{\mathbf{k}_1}{k_1}
g_\lambda(k_1, \omega) - \frac{(\mathbf{k} + \mathbf{q})}{|\mathbf{k} +
\mathbf{q}|}
g_\lambda(|\mathbf{k} + \mathbf{q}|, \omega) - \frac{ (\mathbf{k}_1 -
\mathbf{q})}{|%
\mathbf{k}_1 - \mathbf{q}|} g_\lambda(|\mathbf{k}_1 - \mathbf{q}|, \omega)
\right] %
\Biggr\}\,. \nonumber
\end{eqnarray}
\end{widetext}

For the subsequent discussion we take the limit $\omega=0$ and focus on the solution of the linearized
transport equation in Eq.~(\ref{trans2}) for the function $g$. The following discussion closely follows the presentation given in Ref.~\onlinecite{FSMS}. We can view the right hand side of Eq.~(\ref%
{trans2}) as a linear operator, the so-called collision operator $\mathcal{C}
$, acting on the function $(\mathbf{k}/k)g(k)$. A key
property of $\mathcal{C}$ is that it is Hermitian with respect to the
natural inner product
\begin{equation}
\label{innerprod}
\langle g_{1}|g_{2}\rangle \equiv \sum_\lambda \int \frac{d^{2}k}{(2\pi )^{2}}%
g_{1,\lambda}(k)g_{2,\lambda}(k)\; .
\end{equation}%
This Hermiticity follows \cite{Ziman} from symmetry properties of $R_{1}$
and $R_{2}$ under exchanges between incoming and outgoing momenta, which are
very similar to those used in establishing Boltzmann's H-theorem.

Related to the above properties of the collision operator, we can introduce
a functional $\mathcal{Q}[g]$, such that Eq.~(\ref{trans2}) is equivalent to
finding its stationary point
\begin{equation}
\frac{\delta \mathcal{Q}[g]}{\delta g}=0.  \label{stat}
\end{equation}%
with the
explicit form of the functional given by
\begin{widetext}
\begin{eqnarray}
&&\mathcal{Q}[g]=\frac{(2\pi )}{8v_{F}}\int \frac{d^{2}k}{(2\pi )^{2}}\frac{%
d^{2}k_{1}}{(2\pi )^{2}}\frac{d^{2}q}{(2\pi )^{2}}\Biggl\{  \notag \\
&&~~\frac{\delta (k-k_{1}-|\mathbf{k}+\mathbf{q}|+|\mathbf{k}_{1}-\mathbf{q}%
|)R_{1}(\mathbf{k},\mathbf{k}_{1},\mathbf{q})}{%
(e^{-v_{F}k/T}+1)(e^{v_{F}k_{1}/T}+1)(e^{v_{F}|\mathbf{k}+\mathbf{q}%
|/T}+1)(e^{-v_{F}|\mathbf{k}_{1}-\mathbf{q}|/T}+1)}  \notag \\
&&\times \left[ \frac{\mathbf{k}}{k}g(k )-\frac{\mathbf{k}_{1}}{k_{1}}%
g(k_{1} )-\frac{(\mathbf{k}+\mathbf{q})}{|\mathbf{k}+\mathbf{q}|}g(|%
\mathbf{k}+\mathbf{q}| )+\frac{(\mathbf{k}_{1}-\mathbf{q})}{|\mathbf{k%
}_{1}-\mathbf{q}|}g(|\mathbf{k}_{1}-\mathbf{q}| )\right] ^{2}  \notag
\\
&&+\frac{\delta (k+k_{1}-|\mathbf{k}+\mathbf{q}|-|\mathbf{k}_{1}-\mathbf{q}%
|)R_{2}(\mathbf{k},\mathbf{k}_{1},\mathbf{q})}{%
(e^{-v_{F}k/T}+1)(e^{-v_{F}k_{1}/T}+1)(e^{v_{F}|\mathbf{k}+\mathbf{q}%
|/T}+1)(e^{v_{F}|\mathbf{k}_{1}-\mathbf{q}|/T}+1)}  \notag \\
&&\times \left[ \frac{\mathbf{k}}{k}g(k )+\frac{\mathbf{k}_{1}}{k_{1}}%
g(k_{1})-\frac{(\mathbf{k}+\mathbf{q})}{|\mathbf{k}+\mathbf{q}|}g(|%
\mathbf{k}+\mathbf{q}| )-\frac{(\mathbf{k}_{1}-\mathbf{q})}{|\mathbf{k%
}_{1}-\mathbf{q}|}g(|\mathbf{k}_{1}-\mathbf{q}| )\right] ^{2}\Biggr\}
\notag \\
&&~~~~~~~~~~~~~~~~~~~~~~~-\int \frac{d^{2}k}{(2\pi )^{2}}\frac{g(k)v_{F}/T}{(e^{v_{F}k/T}+1)(e^{-v_{F}k/T}+1)}.
\label{func}
\end{eqnarray}
\end{widetext}

In the previous analysis of a quantum Boltzmann equation for massless Dirac
fermions in two dimensions \cite{ssqhe}, it was noted that the phase space
for scattering of particles was logarithmically divergent in the collinear
limit. For the interaction considered in that paper, the collinear
scattering cross-section vanished, and so this singular phase space density
had no important consequences. The collinear scattering does not vanish for
the present interaction, and so we need to take this logarithmic
divergence seriously~\cite{Kashuba,FSMS}.

The physical origin of the divergent collinear scattering is related to the
linear dispersion which implies that quasiparticles or -holes moving in the same
direction share the same group velocity, independent of their energies. This
leads to a diverging duration of collisions of nearly collinear particles, which
is enhanced due to the low space dimensionality. To the extent that collinear
scattering is very strong, and considering frequencies much smaller than the inelastic scattering rate, we may
expect that quasiparticles and holes that move in the same direction in the
plane will establish a pseudo-equilibrium characterized by an effective chemical
potential and an effective temperature which, however, depends on the
direction of motion.

In linear response the deviations of these effective parameters from the
equilibrium values $\mu$ and $T$ have to vary with ${\bf k}/k\cdot {\bf E}$ for
symmetry reasons. The remaining dominant mode of
the function $g$ will correspond to an effective shift in chemical potential
which translates into
\bea
\label{psimu}
g(k)= \frac{v_F}{T^2}\chi,
\eea
where the prefactor has been chosen so as to make $\chi(\omega)$ dimensionless.
With this Ansatz, which will be confirmed below, it simply remains to determine
the prefactor $\chi$, yielding the leading term in the non-equilibrium
distribution. The whole reasoning presented here is given in much more detail in Refs.~\onlinecite{FSMS,MFS}.
Note that the effective chemical potential shift ranges between $\pm \chi\,\hbar
v_F e E/T$ depending on the direction of motion. Comparing this to the temperature allows
us to estimate the threshold electric field strength, $e E_{\rm lin}=T^2/\hbar
v_F$, below which non-linear effects should remain small.

The Ansatz~\eqref{psimu} can be justified along the lines of Ref.~\onlinecite{FSMS}. The phase space for scattering diverges upon the perpendicular momentum going to zero. However, it turns out that the scattering matrix is not zero for collinear scattering. This implies that the logarithmic divergence for the perpendicular momentum going to zero has to be cut off by higher-order self-energy corrections to the free fermions. This implies that modes which are not zero for the collinear scattering channel have an additional factor $\ln \frac{N}{g_s}$, which in the limit $N \gg 1$ is a large number.

It thus turns out that the diverging phase space for collinear scattering helps to essentially solve the problem exactly in the limit $\ln (N/g_s)\gg 1$.

We conclude that up to
corrections of order $[\ln (N/g_s )]^{-1}$, we can choose $g$ to be of
the form
\begin{equation}
g(k)\approx \frac{v_{F}}{T^{2}}C.  \label{gc}
\end{equation}%
We insert this parameterization into the functional $\mathcal{Q}[g]$ in Eq.~(%
\ref{func}); the solution of the stationarity condition in Eq.~(\ref{stat})
is then equivalent to requiring the vanishing of the derivative with
respect to $C$. We numerically evaluated the integrals in Eq.~(\ref{func})
using an elliptic co-ordinate system to solve the energy conservation
constraint \cite{ssqhe},  and obtained
\begin{equation*}
{\mathcal{Q}}[g]=\frac{1}{T}\frac{\ln {2}}{4\pi }\Biggl[\kappa \frac{g_s^2 T^2}{v_F^2 N} C^{2}-2 C\Biggr]\,,
\end{equation*}%
with $\kappa =1.\,\allowbreak 8448 \times 10^{-4} \pm 1.0\times 10^{-8}$. From the
stationarity condition we then obtain
\begin{equation}
C=\frac{Nv_F^2}{\kappa g_s^{2}T^2}\,.
\end{equation}%
The conductivity can be obtained from $C$ by combining Eqs.~(\ref%
{defj1}), (\ref{paramet}) and (\ref{gc}):
\begin{equation}
\sigma (\omega=0 )=\frac{e^{2}}{h}\frac{v_F^4N^2\ln 2}{\kappa
g_s^{2}\left(k_B T\right)^2}\,,  \label{somega}
\end{equation}%
where we have re-inserted factors of $\hbar $ and $k_B$. This seems to diverge upon lowering temperature, however, we have to keep in mind that at criticality we have $g^\star=\frac{2g_s \Lambda}{v_F^2\pi}=-1$. Identifying the running energy scale $\Lambda$ with the temperature $k_B T$ we end up with 
\begin{eqnarray}\label{Eq:UQCC}
\sigma (\omega=0 )=\frac{e^{2}}{h}\frac{4 N^2\ln 2}{\kappa
\pi^{2}}\;. \label{somega}
\end{eqnarray}
This result is independent of temperature, as we expect for a quantum-critical point in two dimensions. It is worthwhile noting that the prefactor is $\propto N^2$, which is different from the large-N transport analysis of the $O(N)$ rotor model shown in Ref.~\onlinecite{ssbook}, which only has a prefactor of $N$. The difference stems from the fact that in our case all $N$ flavors couple to the electromagnetic field, whereas in the latter case only two components were coupled to the electromagnetic field. So in contrast to the analysis of the transport equations in model I we find that the universal conductivity is finite, however with a numerically huge prefactor. We will now comment on the relationship between the two results.

\subsection{Comparison of results in model I and model II}\label{Sec:Boltzcomp}

We have seen that model I and model II seemingly give contradictory answers to the question whether there is a finite universal conductivity. Whereas in model I we find that a kinematical constraint knocks out scattering from the order parameter fluctuations right at the critical point, in model II we find a finite scattering due to the electron-electron interaction. We will argue in the following that this can actually be interpreted consistently. In order to do so we have to analyze the reason for the kinematical constraint in model I. This discussion can most easily be carried out directly in two dimensions, where for the moment we put all subtleties about controlling the perturbation series aside. If we consider the r.h.s of Eq.~\eqref{Eq:BoltzDirac} we realize that the reason for knocking out scattering processes stems from the fact that the order parameter fluctuations are described by sharp quasiparticles, {\it i.e.}, the spectral function assumes the form of delta peaks. One can now ask in which sense this describes the right physics. In order to do so we can pretend to derive model I from model II via Hubbard-Stratonovich transformation and integrating out electrons. It turns out that the propagator of the order parameter field is actually {\it not} given by 
\begin{eqnarray}\label{Eq:bareboson}
D(\nu_n,{\bf {k}})\propto \frac{1}{\nu_n^2+{\bf{k}}^2}
\end{eqnarray}
but instead by
\begin{eqnarray}\label{Eq:renomboson}
D(\nu_n,{\bf {k}})\propto \frac{1}{\sqrt{\nu_n^2+{\bf{k}}^2}}\;.
\end{eqnarray}
This has severe consequences: while Eq.~\eqref{Eq:bareboson} implies that the bosonic spectral function is sharply peaked, Eq.~\eqref{Eq:renomboson} describes a continuum, which still has its maximum weight at the former resonance. While a lot of the weight still is kinematically blocked, using Eq.~\eqref{Eq:renomboson} in the Boltzmann approach would immediately render the r.h.s. of Eq.~\eqref{Eq:BoltzDirac} finite, albeit small. In that sense the bosonic sector in model I does not faithfully describe the character of the bosonic spectral function in the real two dimensional system. This implies that in reality we expect the real bosonic spectral function to still be peaked around the resonances, but also to have a continuum background. The Boltzmann analysis in model II backs up this statement. Directly in two dimensions we find a finite temperature independent inverse scattering time from interactions with the collective field, albeit with a tiny numerical prefactor. The prefactor being so small can consistently be interpreted as a remnant of the complete kinematical blocking in model I. Our conclusion thus is that the Boltzmann equation as it is presented for model I simply overestimates the kinematical blocking by annihilating the scattering completely. We thus conclude this section by stating that the asymptotic quantum-critical universal conductivity is numerically huge, but independent of temperature, as we expected to find.

\section{Quantum-critical conductivity}\label{Sec:QCRcon}

We will now discuss the implications of the results of Sec.~\ref{Sec:Boltzmann}. We choose to carry out the discussion in the framework of model II for reasons of enhanced clarity. We found that the universal quantum-critical conductivity within model II is a pure number times $e^2/h$. This central result is shown in Eq.~\eqref{Eq:UQCC}. However, as we argued before, the prefactor of the inverse scattering time is numerically tiny, consistent with the complete blocking found in model I. This naturally leads us to consider the influence of the remaining marginally irrelevant interaction, which is given by the long-range Coulomb interaction.

We established that in the clean system at zero chemical potential there are two main sources of current relaxation namely (i) short-range interaction which drives the phase transition and (ii) long-range Coulomb interaction. 

In the following we compare the inverse scattering times due to the two independent processes. 
The discussion is simplified by the fact that we can restrict our discussion of the inverse scattering times to one hydrodynamic mode, see Sec.~\ref{Sec:Boltzmann}. We know that in the framework of the Boltzmann equation collision integrals add. Thus, by virtue of the most relevant relaxation processes for both types of interaction living on the same hydrodynamic mode inverse scattering times also simply add up. This is what is usually called Matthiessen's rule~\cite{Ziman} within the relaxation time approximation, but here it is essentially exact. In the following we again lend from Ref.~\onlinecite{FSMS} and simply take the numerical value of the inverse scattering time on the zero mode and compare it to the one obtained in Eq.~\eqref{Eq:UQCC}. Since for practical purposes the large-N parameter $N$ is a number of order on we can safely drop it from our discussion in the physically relevant situation. Schematically, we can write the collision integral composed of the sum of the two processes as
\begin{eqnarray}
\frac{1}{\tau}=\left(a  +b \alpha^2(T) \right)T\;,
\end{eqnarray}
where 
\begin{eqnarray}\label{Eq:CIflow}
\alpha(T)=\frac{\alpha_0}{1-\frac{\alpha_0}{4} \ln \frac{T}{\Lambda}}\;
\end{eqnarray}
is the running long-range coupling. The perturbation theory of both terms is formally controlled in the small parameter $1/N$. However, there is a big difference: while the first term in brackets is temperature-independent, the second term scales to zero logarithmically upon lowering temperature. This implies that asymptotically in the very low temperature limit the second term vanishes and the first term determines the universal temperature independent quantum-critical conductivity, in agreement with the previous discussion. For higher temperatures, however, the contribution due to long-range Coulomb interaction will win.

We can make a very crude order of magnitude estimate for the temperature $T_{\rm{cr}}$, where the crossover from universal quantum-critical conductivity to the regime, in which conductivity is determined by long-range Coulomb interaction, takes place. We know that in graphene the ultraviolet cutoff is of the order eV, which translates to $\Lambda \approx 10^4 K$. Furthermore, the bare long-range Coulomb interaction is a dimensionless number of order one. We showed in Sec.~\ref{Sec:Boltzmann} that $a\approx 10^{-3}$ and Ref.~\onlinecite{FSMS} found $b\approx 1$. We can estimate the crossover temperature $T_{\rm{cr}}$ as the temperature, where both contributions are comparable in size. This crossover temperature can conservatively be estimated to be of the order $T_{\rm{cr}}\approx 10^{-50}K$. This implies that for all practical purposes the universal quantum-critical conductivity will be masked by the long-range Coulomb interaction. 

Interestingly, this also implies that the quantum-critical conductivity at the semimetal-to-insulator critical point has the same characteristic transport properties as the one in the semimetal, {\it i.e.}, for a gas of hot Dirac fermions interacting via long-range Coulomb interaction. Furthermore, this implies that the conclusions of the hydrodynamic transport relations obtained in a series of papers~\cite{FSMS,MFS,MSF} also hold at the quantum-critical point.

We simply cite the result for the Coulomb interaction limited minimal conductivity which has been analyzed elsewhere~\cite{Kashuba,FSMS,MFS,MSF}. There, a Boltzmann equation was solved directly in two dimensions. This procedure has been presented at length in Ref.~\onlinecite{FSMS} and we only highlight the final result for the d.c. conductivity reading
\begin{eqnarray}\label{Eq:cond}
\sigma = 0.76\frac{e^2}{h}\frac{1}{\alpha(T)^2}\;.
\end{eqnarray}
 Using Eq.~\eqref{Eq:CIflow} in Eq.~\eqref{Eq:cond} we observe that upon lowering the temperature the conductivity diverges as $\alpha$ scales to zero logarithmically. In Eq.~\eqref{Eq:CIflow} $\alpha_0$ is a number of order one and $\Lambda$ constitutes the high energy cutoff set by the lattice. 
 
We end this discussion by saying that we expect this to be the correct minimal conductivity in region II in Fig.~\ref{Fig:phasediag} for all physically relevant situations. This implies that within our approach for realistic temperatures we cannot distinguish the minimal conductivity in region I from region II in Fig.~\ref{Fig:phasediag}.

\section{Discussion and conclusion}\label{Sec:discussion}

In this paper we have investigated quantum-critical transport of electrons on the honeycomb lattice in the vicinity of the semimetal-to-insulator transition. On a technical level, we have used a combination of the semiclassical Boltzmann transport equation with RG and large-N to analyze two related Gross-Neveu type field theories. One of the central results of this paper is given in Eq.~\eqref{Eq:cond} which is the conductivity in the quantum-critical region (region II in Fig.~\ref{Fig:phasediag}) for physically realistic temperatures. The most important characteristic of this minimal d.c. conductivity is that it is set by scattering off the long-range Coulomb interaction, which is a marginally irrelevant coupling at the semimetal-to-insulator transition. The result is surprising since the critical conductivity weakly but explicitly depends upon temperature through the flow of the dimensionless Coulomb interaction parameter~\eqref{Eq:CIflow}. In general, for a Lorentz invariant quantum-critical point in two dimensions described by an interacting fixed point one would expect a conductivity independent of temperature. A further remarkable consequence of our calculation is that to leading order in temperature one cannot distinguish the conductivity in Fig.~\ref{Fig:phasediag} in region I from region II. We furthermore showed that below a very low temperature scale $T_{\rm{cr}}$ a temperature independent universal conductivity takes over, in agreement with expecations.

We finally go back to Fig.~\ref{Fig:phasediag} and analyze possible crossover scenarios. We know that the quantum phase transition microscopically is driven by the interaction parameter $U$. Within the two models we consider this translates into different quantities. Whereas in model I the quantum phase transition is driven by the bosonic mass parameter $t$ in the case of model II the phase transition is driven by the interaction parameter $g$. The crossovers of the system are shown in Fig.~\ref{Fig:phasediag} by doted lines with the associated crossover temperature $T^\star(t)$, which is set by $T^\star(t) \propto t^{\nu z}=t^{1/2}$. 

One conclusion of the preceding discussion is that the universal conductivity as a function of temperature in regions I and II is given by
\begin{eqnarray}
\sigma(T)_{\rm{II},\rm{I}} \propto \frac{e^2}{h}\frac{1}{\alpha^2(T)}\;.
\end{eqnarray}

Another more formal interesting result of our analysis of model I is that it is possible to show that the so-called 'boson-drag' in our setup is exactly zero meaning that an equilibrium treatment of the bosonic degrees of freedom within the fermionic collision integral is justified to leading order in the applied external field ${\bf{E}}$. This statement is more general than that and can be extended to systems with perfect particle-hole symmetry, in which the collective bosonic modes do not break the particle-hole symmetry.

\subsection{Crossover: quantum-critical-to-CDW-insulator}

For the case of the CDW we have an Ising type transition. This implies that due to the absence of Goldstone modes symmetry breaking is possible at finite temperatures and a finite electronic gap is stabilized. For the sake of simplicity we concentrate on a crossover function, which captures the essential physics deep in the quantum-critical regime and in the insulating regime. One such function is of the type
\begin{eqnarray}
\sigma(T)^{CDW}_{\rm{II}\to \rm{III}} \propto \frac{e^2}{h}\frac{1}{\alpha^2(T)+\Delta_{\rm{III}}e^{\frac{\sqrt{t}}{T}}}
\end{eqnarray}
with $\Delta_{\rm{III}}$ being a number. 

\begin{figure}
\includegraphics[width=0.45\textwidth]{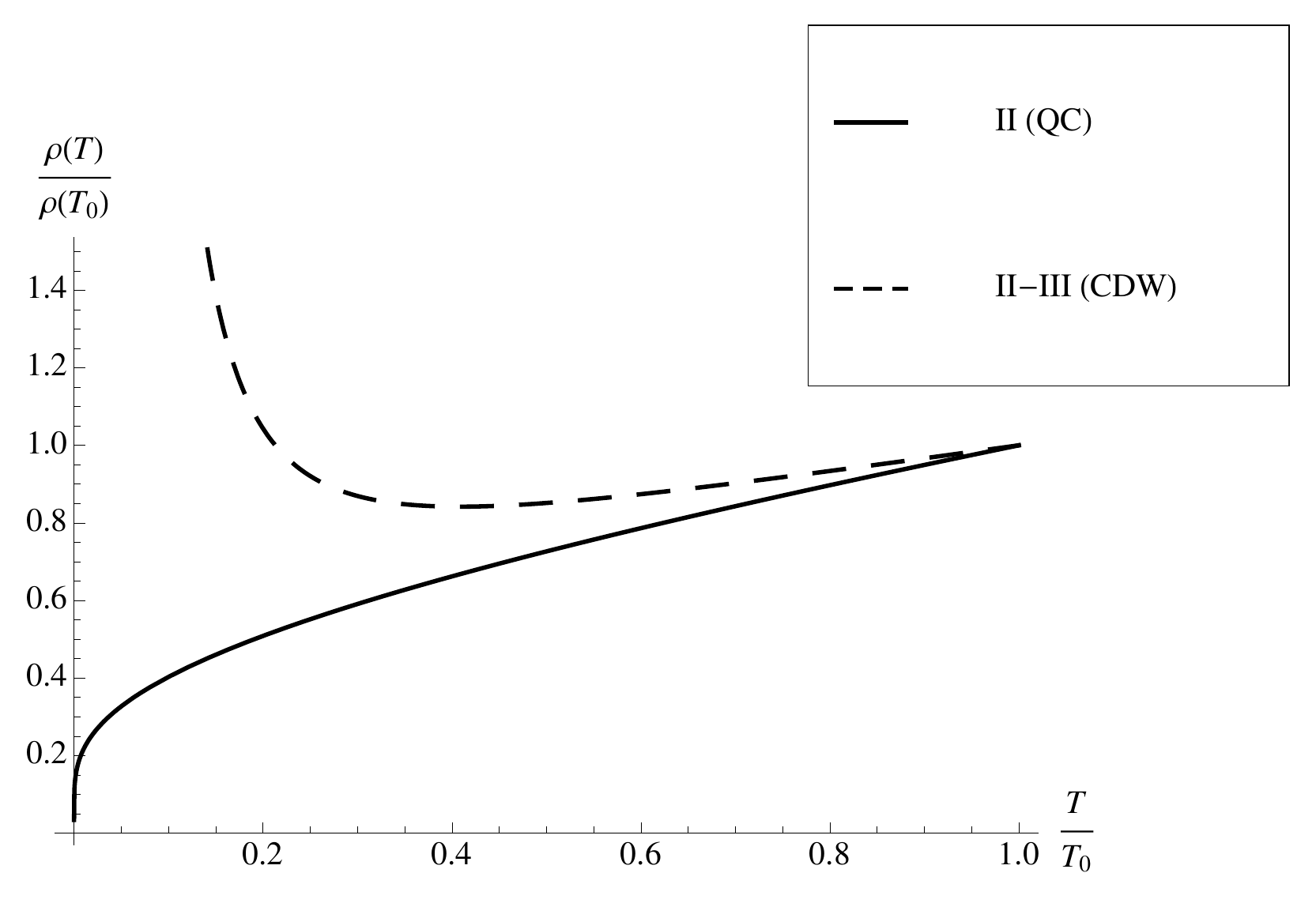}
\caption{Resistivity for regions II (quantum-critical region )and crossover from quantum-critical to CDW called II-III (CDW) as a function of temperature in units of the resistivity at $T=T_0$. The curve for the crossover II-III stays mostly flat over a wide temperature range and eventually takes off exponentially. In the quantum-critical region II the resistivity approaches zero upon lowering temperature in logarithmic manner. The plotted curves are shown for numerical values $\alpha_0=1$ (Eq.~\eqref{Eq:CIflow}), $\Delta_{\rm III}=0.2$, and $t=0.07$.}\label{Fig:resistivity}
\end{figure}

The activated form of the conductivity in the insulating regime III accounts for the fact that we have a finite gap separating the conduction band from a valence band, link in an ordinary semiconductor. A typical crossover curve is shown in Fig.~\ref{Fig:resistivity}, where a comparison of the  quantum-critical resistivity (inverse conductivity right at the critical coupling) is compared with the crossover function for the crossover from region II to region III. The full line in Fig.~\ref{Fig:resistivity} shows the quantum-critical conductivity, whereas the dashed line shows the crossover function. 
 
\subsection{Crossover: quantum-critical-to-SDW-insulator}

\begin{figure}
\includegraphics[width=0.45\textwidth]{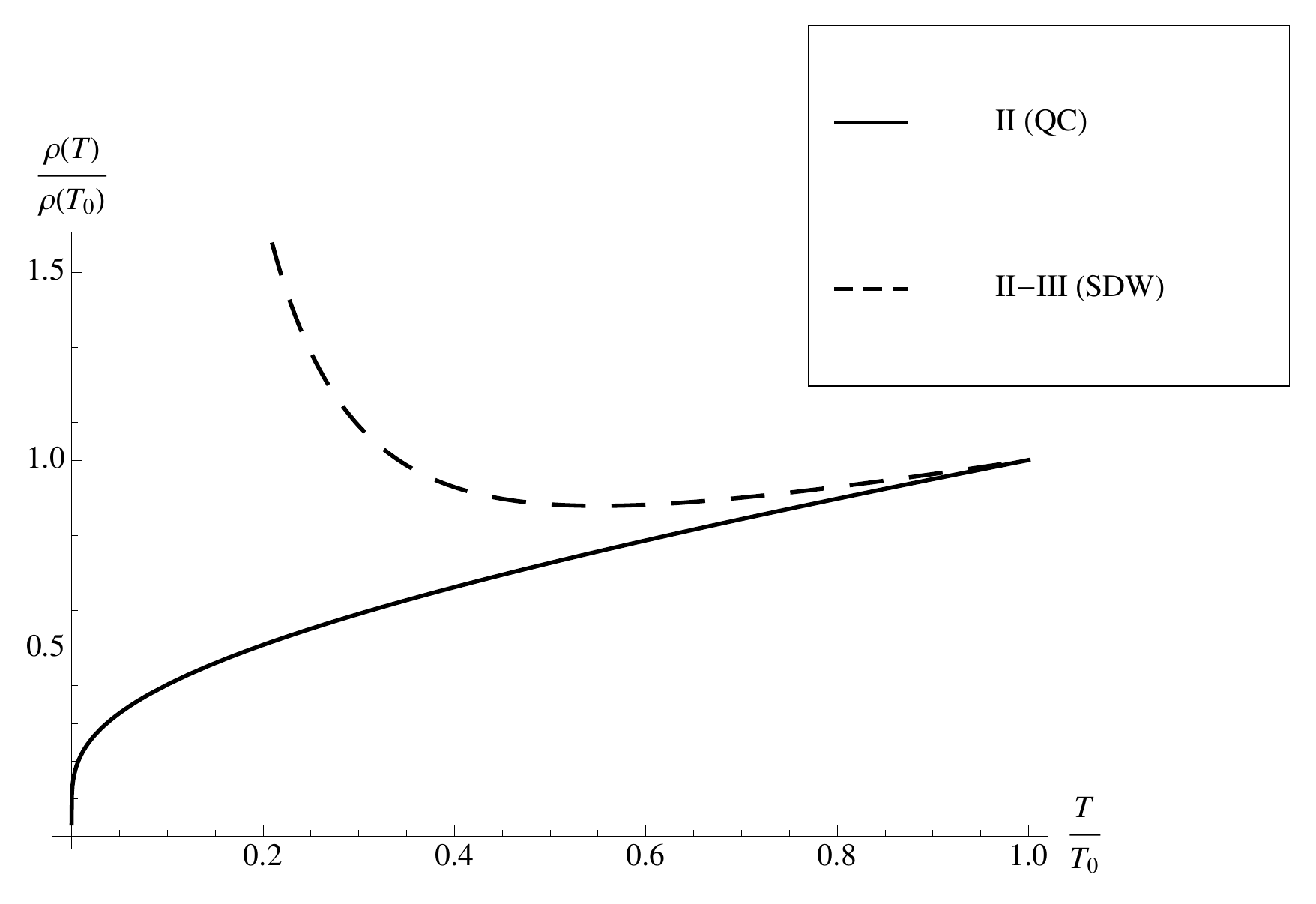}
\caption{Resistivity for regions II (quantum-critical region )and crossover from quantum-critical to SDW called II-III (SDW) as a function of temperature in units of the resistivity at $T=T_0$. The curve for the crossover II-III stays mostly flat over a wide temperature range and eventually takes off as a power-law. In the quantum-critical region II the resistivity approaches zero upon lowering temperature in logarithmic manner. The plotted curves are shown for numerical values $\alpha_0=1$ (Eq.~\eqref{Eq:CIflow}) and $\Delta_{\rm III}=0.1$.}\label{Fig:resistivitySDW}
\end{figure}

In the case of a CDW instability a discrete Ising degree of freedom is spontaneously broken and long-range order at finite temperature is possible.
For the SDW an additional breaking of spin rotation symmetry is required, thus there can be no true long-range order at finite temperature by virtue of the Hohenberg-Mermin-Wagner theorem. This implies that the fermionic quasiparticles are only truly gapped at zero temperature. The question is whether the conductivity can still show signatures of a pseudogap in the conductivity which diverges as temperature is lowered. The role of a finite temperature pseudogap in scaling functions has been analyzed before in the context of the Hertz-Millis theory in two dimensional systems~\cite{Rosch}. In our qualitative consideration, however, we go along the lines of Lee {\it et al.}~\cite{Lee}, who considered fluctuation effects at the Peierls transition in one dimensional systems. It is most convenient to think of the problem at very low temperatures in terms of a spin-fermion model like model I. We concentrate in the following on the low-temperature region which is usually called the renormalized classical regime~\cite{Chubukov}. There, the self energy of the electrons is given by the zero-frequency Matsubara mode of the bosonic propagator, {\it i.e.},
\begin{eqnarray}
D({\bf{q}},\nu_n)\approx \frac{1}{{\bf{q}}^2+\xi^{-2}}
\end{eqnarray}
where $\xi$ is the finite correlation length which diverges upon lowering temperature. In the framework of a Boltzmann theory this implies that the former inelastic scattering of the electrons off the bosonic mode has become elastic and corresponds to the scattering off long-range impurities of the $\frac{1}{r}$-type where $r$ is the distance. In order for this to be true we assume $\frac{\xi^{-1}}{T}=0$ here, which is a non-singular limit in the Boltzmann-equation and corresponds to the "renormalized classical" regime. The temperature behavior of such a type of impurities has been analyzed before~\cite{MFS} and it was found that the associated transport scattering time behaved as 
\begin{eqnarray}
\frac{1}{\tau} \propto \frac{1}{T}\;.
\end{eqnarray}
Again, we assume a crossover function which correctly captures the transport characteristics in the quantum-critical and the renormalized classical regime of the form
\begin{eqnarray}
\sigma(T)^{SDW}_{\rm{II}\to \rm{III}} \propto \frac{e^2}{h}\frac{1}{\alpha^2(T)+\frac{\Delta_{\rm{III}}}{T^2}}\;.
\end{eqnarray}
The respective crossover function is plotted and compared to the quantum-critical resistivity with notation as above in Fig.~\ref{Fig:resistivitySDW}
\subsection{Role of particle-hole symmetry breaking and disorder}

One element we have not yet discussed is the role of particle-hole symmetry breaking and related to this disorder. If the system does not have perfect particle-hole symmetry some rigorous statements, such as the one about vanishing boson-drag and finite conductivity without disorder, cease to be true. In practice, there are several ways to break particle-hole symmetry. One is to allow for a finite chemical potential, which however brings us away from half-filling. Another way is to take into account second neighbor hopping $t'$. If particle-hole symmetry is absent, there is no current relaxation in absence of impurities, since the momentum mode is excited in that case. Also, in that case the boson-drag cannot be neglected any more. However, we still expect our theory to be valid in some limit. The regime of validity of our theory is identical to the one which has been worked out in great detail Ref.~\onlinecite{MFS}. For a finite chemical potential we need $\mu \ll T$ for our theory to hold. For particle-hole symmetry breaking due to next-nearest-neighbor hoppings, such as provided by $t'$~\cite{NetoRMP} we have to require that for the temperature window we are considering that we are still within energies, where the spectrum behaves linearly.

\subsection{Experimental relevance}

Within this paper we have analyzed the resistivity in the vicinity of the semimetal-to-insulator transition on the honeycomb lattice. More specifically, we have studied the crossover from region II to region III in Fig.~\ref{Fig:resistivity} and Fig.~\ref{Fig:resistivitySDW}. 

We now argue that our findings bear some relevance for the understanding of experiments carried out on suspended graphene samples. Remarkably, the curves shown in Fig.~\ref{Fig:resistivity} and Fig.~\ref{Fig:resistivitySDW} mimic the almost constant behavior of the resistivity upon decreasing temperature over a large temperature window followed by an upturn reminiscent of an insulator as observed by Bolotin {\it et. al}~\cite{Bolotin2} on suspended samples (see Ref.~\onlinecite{Bolotin2}, Fig. 4 inset). The samples considered in those experiments exhibit ultrahigh mobilities and enormous mean free paths with respect to impurity scattering (concerning impurity scattering the samples are ballistic with an associated mean free path longer than the sample length). These high mobilities and long elastic mean free paths are obtained after repeated annealing removing impurities from the samples. This makes the inelastic scattering mechanism discussed in this note a viable candidate to set the limiting scattering process for the conductivity. Most strikingly, Bolotin {\it et al.} find that upon lowering temperature the resistivity increases. It is tempting to attribute this behavior to the existence of a quasiparticle gap. 

It is well known that increasing the interaction on the honeycomb lattice, both short- or long-range, can lead to a finite quasiparticle gap in the system. Samples on substrates do not seem to show this tendency, but one has to keep in mind that suspending the sample modifies the dielectric environment and renders the system more strongly interacting. One can thus imagine that suspending graphene samples modifies the dielectric environment strongly enough to drive the system towards strong interactions rendering the charge gap finite. 

At finite densities ($\mu\neq0$) a qualitative explanation of the experimental findings has been given in Ref.~\onlinecite{Trauzettel}. However, here we concentrate on the zero doping case and to the best of my knowledge the data is not fully understood. 

Recently, the data at charge neutrality have been thoroughly reanalyzed by Drut {\it et al.}~\cite{Drut2} who speculated about an underlying, possibly interaction driven gap in the system. The authors determined a couple of phenomenological parameters to fit the data. More specifically, a phenomenological form of the conductivity
\begin{eqnarray}
\sigma=\sigma_q+\sigma_{bg}
\end{eqnarray}
where the subscript q stands for quasiparticle and bg for background was introduced. The background conductivity was set by an unspecified scattering process. Thus we do not speculate about the nature of this part but concentrate on the quasiparticle part. 

For the interpretation and the fitting of $\sigma_{bg}$ the authors considered a gap as well as another unspecific scattering process: concerning the unspecified scattering process, the ratio of the inverse scattering time to temperature, {\it i.e.}< $\frac{1}{\tau T}$ was found to be independent of temperature. Up to logarithmic accuracy this is indeed the case in our scenario, due to the inverse scattering time due to inelastic scattering being given by $\frac{1}{\tau} \propto \alpha^2(T) T$. Furthermore, the authors find that within the relevant temperature range the mean free path associated with this scattering process is on the order of a tenth of micrometers. This is also in agreement with our results. The mean free path corresponding to inelastic scattering due to long-range Coulomb interaction has been estimated before in Ref.~\onlinecite{MFS} and is given by $l_{ee}\approx \frac{2.3 \mu m}{\alpha(T)^2 T[K]} $. In this sense long-range Coulomb interaction is a natural candidate for one of the current relaxation mechanisms needed to explain the measurements faithfully.

Concerning the gap of the quasiparticle, the authors argued that the data is not only compatible with a hard gap but also with a power-law type contribution. The hard gap naturally emerges from the CDW scenario, whereas the pseudogap behavior stems from the SDW scenario. Both mechanisms result in a diverging resistivity at low temperatures. 

Obviously, there are additional other scattering sources, which could play a role. It turns out that phonons due to the high Fermi velocity play a role at rather elevated temperatures ($T>150 K$) and thus are not a candidate to modify the above picture of the transport characteristics~\cite{Adam}. Another candidate would be the corrugation or ripples which are observed in suspended graphene. It turns out that on a formal level ripples share some characteristics with long-range Coulomb interaction. To our knowledge no thorough study of the effect of ripples on the conductivity in suspended graphene exists to date and a detailed analysis is called for~\cite{Adam}.

We thus conclude this discussion by stating that a proximity of free standing graphene to a quantum-critical point for a transition from the semimetal to an insulator described in this note could provide a natural explanation for the (temperature) behavior of the resistivity observed in the experiment of Bolotin {\it et al.}~\cite{Bolotin2} on ultraclean high-mobility samples. This, however, would also imply that signatures of collective symmetry breaking of the SDW or CDW type should be observable in those samples, which to my knowledge has not been attempted. This provides a necessary independent check of the scenario of graphene in vacuum being an excitonic insulator.

\begin{acknowledgments}
During the course of the work I had useful discussions with M. Garst, T. A. L\"ahde, M. M\"uller, A. Rosch, S. Sachdev, J. Schmalian, M. Vojta, and B. Wunsch. 
\end{acknowledgments}

\end{document}